\documentclass[12pt]{article}
\usepackage{times}
\usepackage{geometry}
\usepackage{natbib}
\usepackage{aas_macros}
\usepackage[]{graphicx}
\usepackage{hyperref}
\usepackage{wrapfig}

\usepackage[utf8]{inputenc}
\usepackage{enumitem,amssymb}
\usepackage{ragged2e}
\newlist{thematic}{itemize}{8}
\setlist[thematic]{label=$\square$}
\usepackage{pifont}
\newcommand{\cmark}{\ding{51}}%
\newcommand{\done}{\rlap{$\square$}{\raisebox{2pt}{\large\hspace{1pt}\cmark}}%
\hspace{-2.5pt}}

\hypersetup{
    colorlinks=true,
    linkcolor=blue,
    filecolor=magenta,
    urlcolor=cyan,
}
\begin{document}
\raggedright
\huge
Astro2020 Science White Paper \linebreak

High Angular Resolution Astrophysics: Fundamental Stellar Parameters \linebreak
\normalsize

\noindent \textbf{Thematic Areas:} \hspace*{60pt} $\square$ Planetary Systems \hspace*{10pt} $\square$ Star and Planet Formation \hspace*{20pt}\linebreak
$\square$ Formation and Evolution of Compact Objects \hspace*{31pt} $\square$ Cosmology and Fundamental Physics \linebreak
  $\rlap{$\done$}\square$  Stars and Stellar Evolution \hspace*{1pt} $\square$ Resolved Stellar Populations and their Environments \hspace*{40pt} \linebreak
  $\square$    Galaxy Evolution   \hspace*{45pt} $\square$             Multi-Messenger Astronomy and Astrophysics \hspace*{65pt} \linebreak

\textbf{Principal Author:}

Name:	Gerard van Belle
 \linebreak						
Institution:  Lowell Observatory
 \linebreak
Email: gerard@lowell.edu
 \linebreak
Phone:  928-233-3207
 \linebreak

\textbf{Co-authors:} (names and institutions)\\
 Ellyn Baines, Naval Research Laboratory\\
 Tabetha Boyajian, Louisiana State University\\
 Doug Gies, Georgia State University\\
 Jeremy Jones, Georgia State University\\
 John Monnier, University of Michigan\\
 Ryan Norris, Georgia State University\\
 Rachael Roettenbacher, Yale University\\
 Theo ten Brummelaar, Georgia State University\\
 Kaspar von Braun, Lowell Observatory\\
 Russel White, Georgia State University
  \linebreak

\textbf{Abstract:}
 Direct determination of fundamental stellar parameters has many profound and wide-ranging impacts throughout astrophysics.  These determinations are rooted in high angular resolution observations.  In particular, as long-baseline optical interferometry has matured over the past decade, increasingly large survey samples are serving to empirically ground the basic parameters of these building blocks of the universe.  True imaging and improved parametric fitting are becoming routinely available, an essential component of fully characterizing stars, stellar environments, and planets these stars may host.  

\pagebreak

\section{Fundamental Stellar Parameters}

\begin{wrapfigure}{R}{0.5\textwidth}
\centering
\includegraphics[width=0.98\linewidth, angle=0]{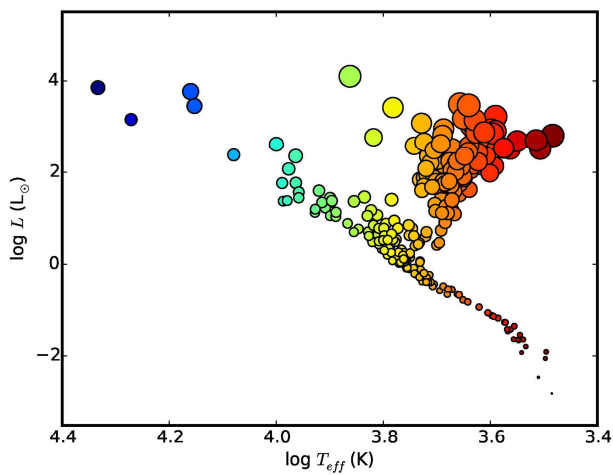}
\caption{\footnotesize The empirical Hertzsprung-Russell diagram, based on {\it direct} angular diameter measurements of nearby stars (up to 150pc) in the literature with random uncertainties $<5\%$ \citep{vonBraun2017arXiv170707405V}.   The size of the points scale with $\log R$, and the colors with $T_{\rm EFF}$.
  \label{fig-HRD}}
\end{wrapfigure}

There are about a trillion stars in the Milky Way. Some are single, some are bound in hierarchical systems, perhaps members of star clusters, but it is readily apparent that the parent star or parent stars of every stellar system dominates the system in multiple, fundamental ways. First, stars will be the principal mass repository in the system, which may additionally contain planets, dust, and gas. Second, stars deeply affect the dynamical, chemical, and evolutionary history of their systems. And third, they provide the main source of energy to their planets, which will be essential for any form of life to form or exist. It is of paramount importance to astronomy to understand how stars function. This insight, however, is very difficult to achieve, in part due to the long lifetimes of stars, the complicated physics governing their structure and evolution, and the paucity of data in the establishment of predictive relations and/or models of stars. Particularly the need to provide constraints to theoretical effort to model stars is addressed by obtaining direct estimates of fundamental stellar parameters (FSPs). These parameters include:
linear radius ($R$);
effective temperature ($T_{\rm EFF}$);
luminosity ($L$);
age ($T$);
mass ($M$);
metallicity, and overall chemical abundance;
rotation, and angular momentum.


\vspace{5pt}
The study of stellar populations and evolution is a cornerstone achievement of 20th century astrophysics \citep[see reviews by][]{Iben1967ARA&A...5..571I,Iben1974ARA&A..12..215I,Iben1983ARA&A..21..271I}.  A central tool of this understanding has been the continued improvement of the Hertzsprung-Russell diagram \citep{Hertzsprung1905WisZP...3..442H,Russell1914PA.....22..275R}.  This diagram, plotting $L$ against $T_{\rm EFF}$, allows for different types of stars, or stars at different stages in their evolution, to be distinguished from each other in a continuous way, which is of particularly high value when populating the diagram with objects with  {\it directly} measured FSPs (Figure \ref{fig-HRD}).

\vspace{5pt}
Such determinations are all part of an ongoing series of tests of the Vogt–Russell theorem\footnote{Although referred to as a `theorem', no formal proof of the Vogt–Russell theorem has been produced.}, which states that the structure of a star, in hydrostatic and thermal equilibrium with all energy derived from nuclear reactions, is uniquely determined by its mass and the distribution of chemical elements throughout its interior.  In particular, this theorem implies the time evolution of stellar luminosity, radius, temperature, and density profiles can all be derived from these initial conditions \citep{Vogt1926AN....226..301V,Russell1926arya.book.....R,Gautschy2015arXiv150408188G,Gautschy2018arXiv181211864G}. Vogt–Russell neglects to include effects of rotation, which also affect a star's structure and evolution, in some cases, significantly \citep{Meynet2016AN....337..827M}.

\section{Interferometry}
Most observational insights of stellar astrophysics come from data involving electromagnetic radiation, such as photometry, spectroscopy, polarimetry, and interferometry.
Here, we concentrate on interferometry with the aim of spatially resolving targets in the diffraction-limited regime.  For all but a handful of the very largest and nearest stars (e.g., Betelgeuse, R Dor), multi-aperture long baseline optical interferometry (LBOI) is required to obtain stellar angular diameters, given their $\sim$milliarcsecond (mas) and smaller size scales.
In their simplest form, interferometric observations result in angular diameters, which can be combined with distance ($d$) and bolometric flux ($F_{\rm BOL}$) to directly determine linear radius and effective temperature.\footnote{Interestingly, luminosity $L$ is {\it not} informed by angular size but instead is contingent upon only $d$ and $F_{\rm BOL}$.  For example, it is possible to have a star of a given luminosity, but it may be hot and small, or cool and large.}  Recently, LBOI has furthermore become able to reveal higher-order surface structure, such as limb and gravity darkening, starspots, and gradients across convective cells (see Figure \ref{fig-zetAnd}).

\vspace{5pt}
Such increasingly accurate and precise observations of fundamental stellar parameters provide excellent tests of and constraints for stellar structure and evolution models, such as \citet{Chabrier1997A&A...327.1039C}, Yonsei-Yale (Y$^2$) \citep{Yi2001ApJS..136..417Y,Demarque2004ApJS..155..667D}, Dartmouth \citep{Dotter2008ApJS..178...89D}, PHOENIX \citep{Husser2013A&A...553A...6H}, and MESA \citep{Paxton2018ApJS..234...34P}.  
For example, 3D simulations of convection in red supergiants with CO$^5$BOLD are being developed to directly interpret and test interferometric observables from LBOI facilities
\citep{Chiavassa2009A&A...506.1351C}.
It is very telling that, for the most abundant type of star in the universe -- the M-dwarf -- our determinations of $T_{\rm EFF}$ and $R$ are {\it significantly} inconsistent with models \citep{Berger2006ApJ...644..475B,Boyajian2012ApJ...757..112B,vonBraun2014MNRAS.438.2413V,Mann2015ApJ...804...64M,Rabus2019MNRAS.484.2674R}.

\section{Work to Date}

Interferometric angular size surveys with 5 or more individual objects in the investigations are listed in Table \ref{tab_previous_surveys}.  The earlier (pre-2000's) surveys of $\geq 4$ mas objects were typically limited in their measurement precision to $\sim$5--15\%; however, given their focus on evolved stars, even measurements at this level of uncertainty were instructive.  More recent surveys have mean uncertainty levels of $\sim$1-5\%, benefiting from more advanced instrumentation at the CHARA, NPOI, and VLTI facilities,
enabling angular diameter work in the 0.4--3 mas regime.

\begin{wrapfigure}{R}{0.55\textwidth}
\centering
\includegraphics[width=0.95\linewidth, angle=0]{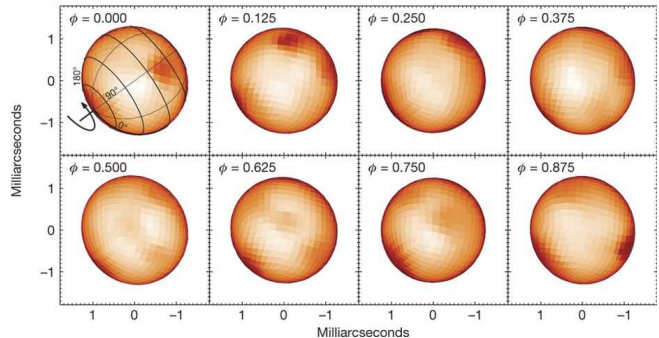}
\caption{\footnotesize Surface mapping of the K1III giant $\zeta$ And by \citet{Roettenbacher2016Natur.533..217R}, revealing limb darkening and surface starspots.
  \label{fig-zetAnd}}
\end{wrapfigure}

\vspace{5pt}
The second major advance in stellar interferometry has been the characterization of stellar surface morphology and imaging of specific stars, primarily enabled by simultanenous observations at multiple baselines. This allowed 2-dimensional characterization of stellar surfaces at multiple spatial scales and was pioneered by \citet{vanbelle2001ApJ...559.1155V}, which measured the non-spherical shape of Altair, and followed by true imaging of Altair's surface in \citet{Monnier2007Sci...317..342M}.  Observations of such rapidly rotating stars has allowed for surface parametrization \citep{Che2011ApJ...732...68C} that includes true rotational velocity, on-sky inclination and orientation, gravity darkening, and polar and equatorial $T_{\rm EFF}$ and $R$ \citep[see review in][]{vanBelle2012A&ARv..20...51V}.

\vspace{5pt}
Extending this reach to less exotic objects has benefited from a rapidly maturing body of image processing and surface mapping routines such as SQUEEZE \citep{Baron2010SPIE.7734E..2IB}, MiRa \citep{Thiebaut2008SPIE.7013E..1IT}, and SURFING \citep{Roettenbacher2016Natur.533..217R}.  These packages have enabled demonstration of the power of LBOI imaging; investigations include the full surface mapping of {$\zeta$} And and {$\sigma$} Gem \citep[Figure \ref{fig-zetAnd};][]{Roettenbacher2016Natur.533..217R,Roettenbacher2017ApJ...849..120R}, and the detection of large granulation cells on $\pi^1$ Gru \citep{Paladini2018Natur.553..310P}.

\begin{table}
\footnotesize
\begin{center}
\begin{tabular}{lll}
\hline
Facility & Survey size & Reference \\
\hline
\multicolumn{3}{c}{Main Sequence Stars} \\
\hline
PTI &  5 Main sequence (MS) stars & \citet{Lane2001ApJ...551L..81L} \\
VLTI-VINCI &  5 Vega-like stars & {\citet{DiFolco2004A&A...426..601D}} \\
VLTI-VINCI/AMBER &  7 low mass stars & {\citet{Demory2009A&A...505..205D}} \\
PTI &  \begin{tabular}{l}40 stars \\(incl. 12 exoplanet hosts) \end{tabular}& \citet{vanBelle2009ApJ...694.1085V} \\
CHARA-Classic &  44 AFG MS stars & \citet{Boyajian2012ApJ...746..101B} \\
CHARA-Classic &  22 KM stars & \citet{Boyajian2012ApJ...757..112B} \\
CHARA-PAVO & 5 MS stars & \citet{Huber2012ApJ...760...32H} \\
CHARA-Classic &  \begin{tabular}{l} 23 A-K stars \\(incl. 5 exoplanet hosts)\end{tabular} & \citet{Boyajian2013ApJ...771...40B} \\
CHARA-PAVO & 7 MS, 3 giant stars & \citet{Maestro2013MNRAS.434.1321M} \\
CHARA-Classic &  11 GKM exoplanet hosts & \citet{vonBraun2014MNRAS.438.2413V} \\
\begin{tabular}{l} CHARA-Classic, \\CLIMB, PAVO\end{tabular} &  7 A-type stars & \citet{Jones2015ApJ...813...58J} \\
CHARA-PAVO & 5 exoplanet hosts & \citet{White2018MNRAS.477.4403W} \\
VLTI-PIONIER & 13 M dwarfs & \citet{Rabus2019MNRAS.484.2674R} \\
CHARA-PAVO & 6 O-type stars & \citet{Gordon2018ApJ...869...37G} \\
CHARA-PAVO & 25 B-type stars & \citet{Gordon2019ApJ...873...91G} \\
\hline
\multicolumn{3}{c}{Giants} \\
\hline
Mark III &  24 giants & \citet{Hutter1989ApJ...340.1103H} \\
Mark III &  12 giants & \citet{Mozurkewich1991AJ....101.2207M} \\
IOTA &  37 giants & \citet{Dyck1996AJ....111.1705D} \\
IOTA &  74 giants & \citet{Dyck1998AJ....116..981D} \\
PTI &  69 giants/supergiants & \citet{vanBelle1999AJ....117..521V} \\
NPOI &  50 giants & \citet{Nordgren1999AJ....118.3032N} \\
NPOI &  41 giants & \citet{Nordgren2001AJ....122.2707N} \\
Mark III &  85 giants & \citet{Mozurkewich2003AJ....126.2502M} \\
CHARA-Classic &  25 K giants & \citet{Baines2010ApJ...710.1365B} \\
NPOI &  69 giants, 18 additional stars & \citet{Baines2018AJ....155...30B} \\
\hline
\multicolumn{3}{c}{Other Evolved Stars} \\
\hline
IOTA &  15 carbon stars & \citet{Dyck1996AJ....112..294D} \\
IOTA &  18 O-rich Miras & \citet{vanBelle1996AJ....112.2147V} \\
IOTA &  \begin{tabular}{l} 9 carbon/S-type Miras, \\4 non-Mira S-type \end{tabular} & \citet{vanBelle1997AJ....114.2150V} \\
IOTA &  22 O-rich Miras & \citet{vanBelle2002AJ....124.1706V} \\
VLTI-VINCI &  14 Miras & {\citet{Richichi2003Ap&SS.286..219R}} \\
VLTI-VINCI &  7 Cepheids & {\citet{Kervella2004A&A...416..941K}} \\
PTI &  74 supergiants & \citet{vanBelle2009MNRAS.394.1925V} \\
PTI &  5 carbon stars & {\citet{Paladini2011A&A...533A..27P}} \\
PTI &  41 carbon stars & \citet{vanBelle2013ApJ...775...45V} \\
VLTI-PIONIER &  9 Cepheids & {\citet{Breitfelder2016A&A...587A.117B}} \\
\hline
\end{tabular}
\caption{Long-baseline optical interferometry diameter surveys of 5 or more stars. See Table 3.1 in \citet{vonBraun2017arXiv170707405V} for a detailed and more comprehensive list of stars with interferometrically determined radii; see also Figure \ref{fig-HRD}.}\label{tab_previous_surveys}
\end{center}
\end{table}
\section{Impacts}

The impact of interferometric measurements have a wide and profound reach within astronomy.

\vspace{5pt}
The large and increasing number of interferometrically characterized stars allows for the calibration of relations of observable quantities with stellar astrophysical parameters, which are almost entirely direct in the sense that no stellar or spectral modeling is necessary. In return, these relations are applicable to stars unavailable to interferometry. Work by \citet{2014AJ....147...47B} and \citet{2018MNRAS.473.3608A} relates stellar angular diameter to broadband, photometric stellar color across a broad range of luminosity classes and spectral types. Studies relating stellar $T_{\rm EFF}$ to broadband color include \citet{Boyajian2013ApJ...771...40B} and \citet{Mann2015ApJ...804...64M}. \citet{Casagrande2010A&A...512A..54C} establishes the well-known infrared flux method based on LBOI results. Studies like \citet{Dyck1996AJ....111.1705D,Dyck1996AJ....112..294D,Dyck1998AJ....116..981D,vanBelle1999AJ....117..521V} calibrate $T_{\rm EFF}$ scales for larger datasets like the {\it APOGEE} spectroscopic survey, and work by \citet{Creevey2012A&A...545A..17C,Creevey2019arXiv190202657C} addresses the metal-poor stellar populations.
The impressive results from the {\it Gaia} mission also greatly benefit from empirical calibration via LBOI, e.g., the Cepheid distance scale \citep{Kervella2018IAUS..330..305K} and double-lined spectroscopic binary masses \citep{Halbwachs2016MNRAS.455.3303H}.  
Conversely, photometric calibration, which in many cases has its roots tied to an empirical reference to Vega, is influenced by the observations of that star as a rapid rotator by LBOI \citep{Peterson2006ApJ...636.1087P,Aufdenberg2006ApJ...645..664A,Gray2007ASPC..364..305G} and its debris disk \citep{Ciardi2001ApJ...559.1147C}.

\vspace{5pt}
Beyond the regime of empirical calibration, high spatial resolution observations provide deep insight into the nature of stars and their underlying physics, especially when combined with other data products. For instance, a deep synergy exists between LBOI observations and asteroseismology, with impacts on the results from TESS, Gaia, and Kepler/K2 \citep{Huber2016arXiv160407442H}.  Detailed characterizations of individual stars that include radii, masses, ages, and $T_{\rm EFF}$'s, are possible through the union of these two techniques, often at percent-level precisions. Similarly, combining stellar angular diameter with spectroscopic and photometric time-series data of transiting exoplanet systems allows for the full and direct characterization of stellar and planetary astrophysical and orbital parameters \citep{2012ApJ...753..171V}.
Connecting stellar surface morphology to the underlying physics -- such as flaring and starspots \citep{Roettenbacher2018ApJ...868....3R} -- is enabled by LBOI stellar surface imaging; astrometric results may be affected by large scale object asymmetries \citep{Cruzalebes2015MNRAS.446.3277C,Chiavassa2018A&A...617L...1C}.
Precision radial velocity (RV) machines such as EXPRES and NEID are pushing towards cm/s levels of precision using techniques such as `spectro-perfectionism' \citep{Bolton2010PASP..122..248B,Petersburg2019AAS...23322501P}.  Combining contemporaneous surface morphology with Doppler RV measurements holds the potential to be a part of the solution in breaking through the $\sim$20 cm/s barrier that limits RV detection of Earth-like objects.
Other serendipitious discoveries await at the LBOI regime of imaging, such as direct observations of diameters and shapes of expanding fireballs of novae, demonstrated with observations of Nova Del 2013 \citep{Schaefer2014Natur.515..234S}, which has provided a window into dynamics and morphologies of these runaway thermonuclear explosions at milliarcsecond scales.
Evolutionary theories of low-mass stars can be explored by studying their final states: direct diameters of the nearest white dwarf stars can explore radius discrepancies of models \citep{Joyce2018MNRAS.479.1612J,Joyce2018MNRAS.481.2361J,Romero2019MNRAS.484.2711R}. Stellar surface imaging could also be used to reveal a transit movie of an exoplanet hosting star \citep{vanbelle2008PASP..120..617V}.

\section{Observational Horizons}

%

Extending the impact of these empirical measurements to a larger sample of stars is crucial to keeping our understanding of stellar astrophysics firmly rooted on an observational basis. As an example, Figure \ref{fig-HRD} illustrates our observational frontiers: more fully populating the main sequence at both the high mass and low mass regimes, is an obvious goal.  Throughout the HR diagram, better sampling is necessary for disentangling second order effects, such as metallicity and rotation.  This will require: (a) {\bf greater angular resolution}, with diameters down to $\sim$10 microarcseconds; and (b) {\bf greater sensitivity}, to $V$ and $K$ magnitudes of at least 12.  Doing so while maintaining uncertainty limits at 2\% or better will ensure the results are relevant and constraining for state-of-the-art stellar models.

\vspace{5pt}

As evidenced by Figure \ref{fig-zetAnd}, LBOI is capable of producing stellar surface imaging; providing this capability routinely to the community would open up entire new avenues of understanding stellar physics in much greater detail than currently possible.  Improving the supporting measurements of distance, $F_{\rm BOL}$, radial velocity, etc., are required to fully capitalize upon advances in high spatial resolution observations. Limb-darkening laws and contrasts due to stellar spots could be empirically calibrated rather than predicted.
The applicability of a thus increased number of direct stellar characterizations plus the resulting calibrations of predictive, semi-empirical relations, is ubiquitous. One example is the rapidly evolving field of exoplanets, especially in the Kepler/K2/TESS age, which has produced thousands of transiting exoplanet candidates and continues to do so. Characterizing these planets is of vital importance to understand their formation and evolution theories, assessing planetary habitability, plus statistical studies such as planetary frequencies and astrophysics as functions of stellar parameters, to name a few specific examples. ``You only understand the exoplanet as well as you understand its parent star." The lingering discrepancy between models and empirical stellar radii and effective temperatures, especially in the low-mass regime where most stars and indeed most planets are located, will continue to bias many statistical statements based on the results from these and other surveys.






%
%
%

\pagebreak
\bibliography{journal-references}

\end{document}